\documentclass[aps,prl,reprint,groupedaddress,showpacs,amsmath,amssymb,floatfix]{revtex4-1}

\usepackage{graphicx}
\usepackage{bm}
\usepackage{color}
\usepackage[normalem]{ulem}
\usepackage{cleveref}
\usepackage{amsmath}
\usepackage{braket}
\usepackage{xspace}

\newcommand{\wse}{WSe$_2$\xspace}
\newcommand{\moire}{moir\'e\xspace}
\newcommand{\Moire}{Moir\'e\xspace}

\newcommand{\rucl}{$\alpha$-RuCl$_3$\xspace}
\newcommand{\sio}{SiO$_2$\xspace}

\newcommand{\dens}{$\times10^{12}$ cm$^{-2}$\xspace}

\newcommand{\twse}{tWSe$_{2}$\xspace}

\newcommand{\rxy}{$R_{xy}$\xspace}

\begin{document}

\title{Superconductivity in twisted bilayer WSe$_2$}

\author{Y. Guo$^{1}$}
\author{J. Pack$^{1}$}
\author{J. Swann$^{1}$}
\author{L. Holtzman$^{2}$}
\author{M. Cothrine$^{3}$}
\author{K. Watanabe$^{4}$}
\author{T. Taniguchi$^{5}$}
\author{D.G. Mandrus$^{3,6}$}
\author{K. Barmak$^{2}$}
\author{J. Hone$^{7}$}
\author{A.J. Millis$^{1,8}$}
\author{A. Pasupathy$^{1,9}$}
\author{C.R. Dean$^{1}$}
\email{Email: cd2478@columbia.edu}

\affiliation{$^{1}$Department of Physics, Columbia University, New York, NY 10027, USA}
\affiliation{$^{2}$Department of Applied Physics and Applied Mathematics, Columbia University, New York, New York 10027, United States}
\affiliation{$^{3}$Department of Materials Science and Engineering, University of Tennessee, Knoxville, Tennessee 37996, United States}
\affiliation{$^{4}$Research Center for Electronic and Optical Materials, National Institute for Materials Science, 1-1 Namiki, Tsukuba 305-0044, Japan}
\affiliation{$^{5}$Research Center for Materials Nanoarchitectonics, National Institute for Materials Science,  1-1 Namiki, Tsukuba 305-0044, Japan}
\affiliation{$^{6}$Materials Science and Technology Division, Oak Ridge National Laboratory, Oak Ridge, Tennessee 37831, United States}
\affiliation{$^{7}$Department of Mechanical Engineering, Columbia University, New York, NY 10027, USA}
\affiliation{$^{8}$Center for Computational Quantum Physics, Flatiron Institute; New York, USA}
\affiliation{$^{9}$Condensed Matter Physics and Materials Science Division, Brookhaven National Laboratory; Upton, USA.}

\date{\today}

\maketitle

\textbf{The discovery of superconductivity in twisted bilayer and twisted trilayer graphene\cite{caoCorrelatedInsulatorBehaviour2018,yankowitzTuningSuperconductivityTwisted2019,luSuperconductorsOrbitalMagnets2019,parkTunableStronglyCoupled2021,haoElectricFieldTunable2021} has generated tremendous interest.  The key feature of these systems is an interplay between interlayer coupling and a \moire superlattice that gives rise to low-energy flat bands with strong correlations\cite{bistritzerMoireBandsTwisted2011}.  Flat bands can also be induced by \moire patterns in lattice-mismatched and or twisted heterostructures of other two-dimensional materials such as transition metal dichalcogenides (TMDs)\cite{wuHubbardModelPhysics2018,wuTopologicalInsulatorsTwisted2019,naikUltraflatbandsShearSolitons2018,ruiz-tijerinaInterlayerHybridizationMoir2019,schradeSpinvalleyDensityWave2019}.
Although a wide range of correlated phenomenon have indeed been observed in the \moire TMDs\cite{wangCorrelatedElectronicPhases2020,tangSimulationHubbardModel2020,xuTunableBilayerHubbard2022,andersonProgrammingCorrelatedMagnetic2023,reganMottGeneralizedWigner2020,liImagingTwodimensionalGeneralized2021,xuCorrelatedInsulatingStates2020,liQuantumAnomalousHall2021,caiSignaturesFractionalQuantum2023,zengIntegerFractionalChern2023,parkObservationFractionallyQuantized2023,fouttyMappingTwisttunedMultiband2024,xuObservationIntegerFractional2023}, robust demonstration of superconductivity has remained absent\cite{wangCorrelatedElectronicPhases2020}.  
Here we report superconductivity in 5 degree twisted bilayer WSe$_2$ (\twse) with a maximum critical temperature of 426~mK. 
 The superconducting state appears in a limited region of displacement field and density that is adjacent to a metallic state with Fermi surface reconstruction believed to arise from antiferromagnetic order\cite{ghiottoStonerInstabilitiesIsing2024}. A sharp boundary is observed between the superconducting and magnetic phases at low temperature, reminiscent of spin-fluctuation mediated superconductivity\cite{mathurMagneticallyMediatedSuperconductivity1998}. 
Our results establish that \moire flat-band superconductivity extends beyond graphene structures.  Material properties that are absent in graphene but intrinsic among the TMDs such as a native band gap,  large spin-orbit coupling, spin-valley locking, and magnetism 
offer the possibility to access a broader superconducting parameter space than graphene-only structures.}

\noindent\textbf{Introduction}

\noindent
Flat bands in 2D heterostructures have been a focus of intense study as a way to achieve strongly correlated electronic states.
In twisted graphene systems, superconductivity \cite{caoUnconventionalSuperconductivityMagicangle2018,yankowitzTuningSuperconductivityTwisted2019,parkTunableStronglyCoupled2021,haoElectricFieldTunable2021}, magnetic ordering \cite{sharpeEmergentFerromagnetismThreequarters2019} and topological Chern bands \cite{serlinIntrinsicQuantizedAnomalous2020} have been observed around commensurate fillings of the \moire band. Superconductivity has also been observed in crystalline graphene, such as rhomboherdral trilayer graphene\cite{zhouSuperconductivityRhombohedralTrilayer2021} and Bernal bilayer graphene\cite{zhouIsospinMagnetismSpinpolarized2022, zhangEnhancedSuperconductivitySpin2023,liTunableSuperconductivityElectron2024}, where high density of states is generated at Van Hove singularities (VHs) located at the boundary of Fermi pocket topology transitions. It remains a question of significant debate whether superconductivity in the graphene systems is a conventional BCS type,  mediated by phonons\cite{wuTheoryPhononMediatedSuperconductivity2018,lianTwistedBilayerGraphene2019,ochiPossibleCorrelatedInsulating2018,chouAcousticphononmediatedSuperconductivityBernal2022}, or unconventional, mediated by another mechanism such as spin fluctuations\cite{xuTopologicalSuperconductivityTwisted2018,isobeUnconventionalSuperconductivityDensity2018,guoPairingSymmetryInteracting2018,rayWannierPairsSuperconducting2019,chichinadzeNematicSuperconductivityTwisted2020,chatterjeeIntervalleyCoherentOrder2022}, or even contains both types. 
 

\begin{figure*}
\includegraphics[width=\linewidth]{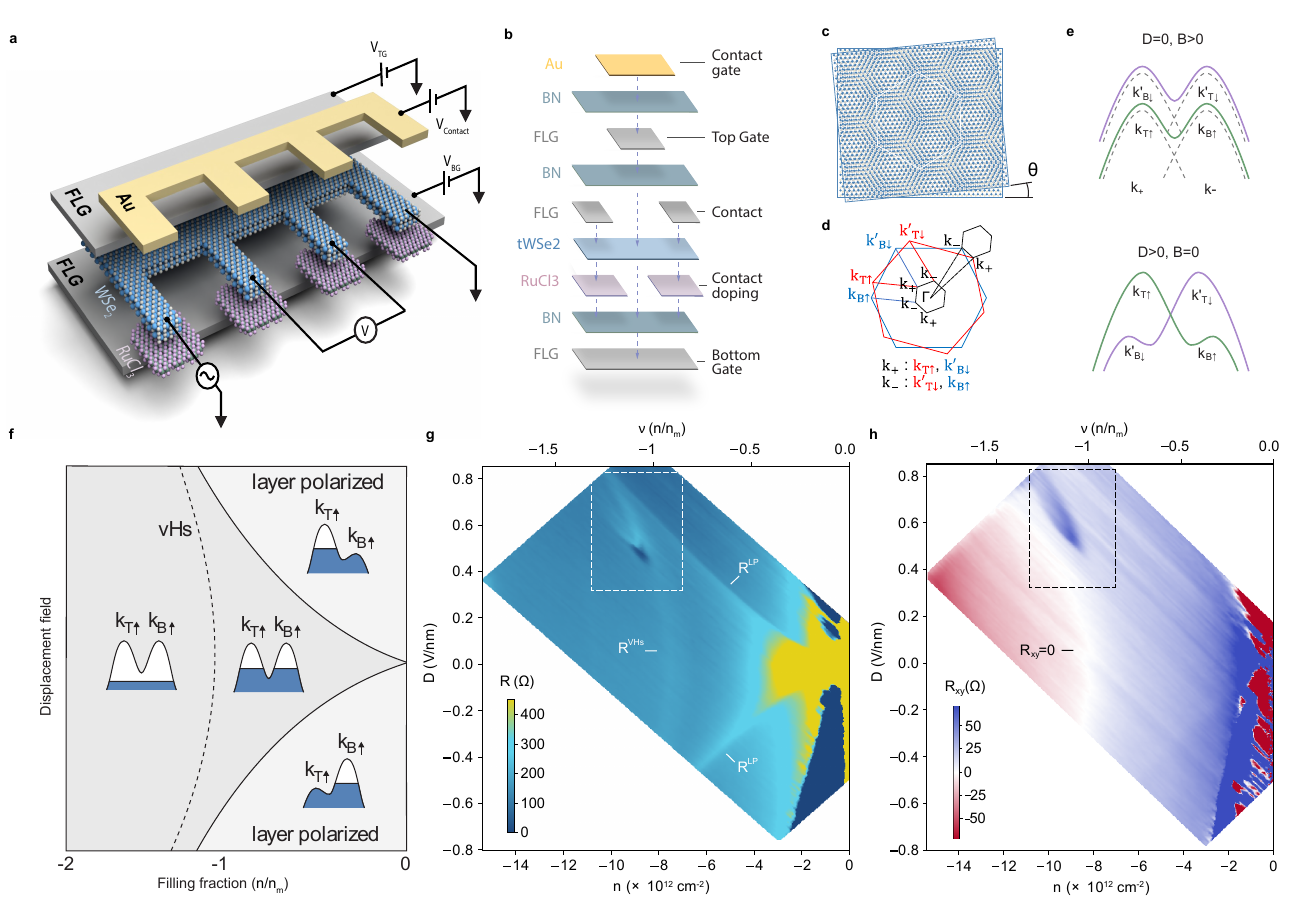}
\caption{{\bf{Electronic band structure and the superconducting pocket}}  \textbf{a},\textbf{b} Schematic illustration of the t\wse device structure. Graphite contacts together with charge transfer doping by  \rucl enables Ohmic contact (see text). \textbf{c}, Illustration of real space \moire pattern between twisted \wse layers. \textbf{d}, Brillouin zones of the top (red) and bottom (blue) layer. The spin states near the $k$ valleys are marked. The small black hexagon shows the \moire mini Brillouin zone. Both the spin up (down) valley of the top layer and the spin down (up) valley of the bottom layer folds onto the $k_{+}$ ($k_{-}$) valley of the \moire mini Brillouin zone. \textbf{e}, Illustration of the low energy \twse band structure.  The spin-valley branches hybridize across the layers as shown. 
 Magnetic field splits the bands via Zeeman coupling (top) where as displacement field distorts each band by layer polarizing.  \textbf{f}, Schematic phase diagram of of the band structure and Fermi energy for a single isospin branch versus displacment field and carrier density \textbf{g}.\textbf{h}, Longitudinal resistance $R$ (\textbf{g}) at zero magnetic field and Hall resistance \rxy (\textbf{h}) at 200~mT versus density and displacement field. The densities are displayed in negative values to represent hole doping. Labels identify transport features that coincide with the solid and dashed boundaries in \textbf{f} (see text). Filling fraction is defined as number of holes per \moire unit cell.
}
\end{figure*} 

\Moire-patterned transition metal dichalcogenides (TMDs) also host low energy flat bands\cite{wuHubbardModelPhysics2018}  and have proven to be a highly tunable system for hosting exotic correlated phases, such as Mott insulators \cite{wangCorrelatedElectronicPhases2020,tangSimulationHubbardModel2020}, generalized Wigner crystal states \cite{reganMottGeneralizedWigner2020,liImagingTwodimensionalGeneralized2021,xuCorrelatedInsulatingStates2020}, integer and fractional quantum anomalous Hall states \cite{liQuantumAnomalousHall2021,caiSignaturesFractionalQuantum2023,zengIntegerFractionalChern2023,parkObservationFractionallyQuantized2023,fouttyMappingTwisttunedMultiband2024,xuObservationIntegerFractional2023} and magnetic ordering \cite{xuTunableBilayerHubbard2022,andersonProgrammingCorrelatedMagnetic2023}. Theoretical studies have suggested that superconductivity should also be present in this system \cite{venderleyDensityMatrixRenormalization2019,slagleChargeTransferExcitations2020,hsuSpinvalleyLockedInstabilities2021,schradeNematicChiralTopological2021,belangerSuperconductivityTwistedBilayer2022,schererChiralSuperconductivityEnhanced2022,crepelTopologicalSuperconductivityDoped2023,chenSingletTripletPair2023,holleisIsingSuperconductivityNematicity2023,kleblCompetitionDensityWaves2023,wuPairdensitywaveChiralSuperconductivity2023,zhouChiralNodalSuperconductors2023,zegrodnikMixedSinglettripletSuperconducting2023,xieOrbitalFuldeFerrellPairing2023,akbarTopologicalSuperconductivityMixed2024,kielyContinuousWignerMottTransition2023}, and with the possibility of exotic character. However, definitive experimental observation remains absent. This raises the question as to whether superconductivity is a universal feature in flat-band 2D systems, or if there is some unique feature in the graphene-based structures that plays a key role. 

Here we report observation of superconductivity in bilayer \wse fabricated with a 5.0$^o$ twist angle. At zero displacement field we do not observe evidence of correlated behaviour, which may be expected since at this twist angle the \moire band is nearly ten times wider\cite{wangCorrelatedElectronicPhases2020} than that of magic angle twisted bilayer graphene, and can thus be expected to host weaker correlations.  However, when an applied displacement field is tuned such that the Van Hove singularity (VHs) of the  \moire band shifts close to half filling (one hole per \moire unit cell), a local region of robust superconductivity emerges adjacent to a resistive metallic state that we associate with anti-ferromagnetic ordering\cite{ghiottoStonerInstabilitiesIsing2024}. Analysis of the SC-AFM phase diagram suggests the superconductivity may be stabilized by spin fluctuations, analogous to similar behaviour seen in heavy Fermion systems\cite{mathurMagneticallyMediatedSuperconductivity1998}.  Since magnetism and superconductivity are both favoured by a high density of electronic states, they often emerge near Van Hove singularities (VHs) and can exhibit either competition or cooperation. The case where magnetic fluctuations mediate superconductivity is particularly interesting since some of the same fundamental electronic interaction determine the ground state\cite{mathurMagneticallyMediatedSuperconductivity1998}.


\vspace{\baselineskip}
\noindent\textbf{Tunable flat bands}

\noindent
Fig. 1a-b shows a schematic illustration of the device structure. Ohmic contacts were realized by using a combination of graphite as the contact metal and RuCl$_{3}$ as a charge-transfer dopant in the contact region\cite{packChargetransferContactHighMobility2023}.  The device was assembled by the dry transfer method\cite{wangOneDimensionalElectricalContact2013} using exfoliated hBN as dielectric spacers (Fig. 1b). The ``cut and stack'' technique\cite{kimVanWaalsHeterostructures2016} was employed to realize an AA-stacked t\wse with twist angle measured to be $5.0^{o}$ (see SI). The \wse layers are derived from high-purity crystals grown by a two-step flux synthesis method\cite{liuTwoStepFluxSynthesis2023}. We defined a three gate structure where ``top'' and ``bottom'' graphite gates allow tuning of the channel density and displacement field, with a third metal gate used to maintain a low electrostatic barrier in the contact region (see methods for details of the device fabrication process).  The contact doping scheme enables reliable contact under hole doping only, and therefore we restrict our discussion to studies of the valence band. The contact resistance, estimated from the two-terminal response, is $\sim$ 12 k$\Omega$ for most of the gate range, rising to 50 k$\Omega$ at density of 0.95\dens (See SI).

In  monolayer \wse, owing to the strong spin-orbit coupling, the spin and valley degrees of freedom are locked, creating a combined spin-valley isospin with opposite spin polarization in each of the the $k$ and $k'$ valleys. In twisted bilayer \wse, the lattice orientation mismatch of the rotated layers results in a \moire superlattice in real space. 
At a twist angle of $5.0^{o}$, the \moire wavelength is 3.76nm\cite{yankowitzEmergenceSuperlatticeDirac2012}. In reciprocal space, a corresponding \moire mini Brillouin zone is formed (Fig. 1d). The $k\downarrow$ ($k'\uparrow$) valley of the top layer and the $k'\uparrow$ ($k\downarrow$) valley of the down layer in the original Brillouin zone both fold onto the $k+$ ($k-$) valley in the \moire mini Brillouin zone. Therefore, each valley of the \moire superlattice contains both spin flavours.

At small twist angles, interlayer hopping causes hybridization between the layer bands, leaving a saddle point where they intersect (Fig. 1e).  The reconstructed bands maintain spin and valley locking, and thus retain a two-fold combined isospin flavour.  
Band structure calculations\cite{wangCorrelatedElectronicPhases2020} indicate that at 5$^{o}$ twist angle, the band width is approximately $w=93$~meV at zero displacement field, whereas the Coulomb energy is $U=109$~meV.  This gives a ratio of $U/t\sim1$, indicating a regime of moderate correlations.  
Applying a displacement field distorts each band, varying its width, causing the VHs position to shift and increasing the DOS at the VHs (see SI)\cite{wangCorrelatedElectronicPhases2020}. At sufficiently large bias and low doping, the displacement field can cause the carriers to polarize to one layer. A cartoon schematic of the VHs position and layer polarization boundaries as a function of displacement field and band filling is shown in Fig. 1f.  The effective low energy band structure for just one of the isospin flavours is shown in each region (see SI for detailed calculation of band structure versus displacement field at the relevant angle of $\sim5^{o}$).

Fig. 1g  shows the longitudinal resistance, $R$ as a function of density and displacement field, measured at $B=0$~T and cryostat temperature of $T=33$~mK. 
The conversion from applied gate potentials to density, displacement field, and band filling fraction was determined from Landau level trajectories (see SI). We define the filling fraction as the number of  holes per \moire unit cell,  $\nu=n/n_{m}$, where full band filling corresponds to $\nu=2$ owing to the twofold isospin degeneracy.   The diverging $R$ seen at low density (yellow in the chosen color scale of Fig. 1g) corresponds to the Fermi level nearing the top of the valence band. The layer polarization boundaries and VHs position are also evident, appearing as local peaks in $R$ (light blue trajectories in Fig 1g). The layer polarization boundaries, labelled $R^{LP}$ in Fig. 1g, were confirmed by measurement in the quantum Hall regime where there is a corresponding change in the LL crossing behaviour (see SI). The behaviour of the resistive peak at the approximate VHs position, labelled $R^{VHs}$ in Fig. 1g, is consistent with bandstructure calculations, which predict that the VHs is located slightly below half band filling at $D=0$, and passes through half-filling at approximately $D\sim 0.4$~V/nm\cite{crepelBridgingSmallLarge2024}.(See SI) Fig. 1h shows the density and displacement field dependence of the Hall resistance, $R_{xy}$, measured at $B=0.2$~T.  $R_{xy}$ changes sign along a similar trajectory (labelled $R_{xy}=0$) as the $R^{VHs}$ peak, but with a slight offset.

The temperature dependence of the resistivity indicates that the material is metallic at all doping levels and displacement fields, in contrast to previous studies of devices with smaller twist angles which showed small regions of fully insulating (but modest bandgap) behavior\cite{wangCorrelatedElectronicPhases2020,ghiottoStonerInstabilitiesIsing2024}.  We attribute the difference to the larger twist angle (implying larger bandwidth) and reduced BN thickness (implying potentially more screening by the gate). Under finite displacement field, approximately when the VHs crosses half-filling (dashed squares in Figs. 1g,h), we observe a local feature with anomalous behavior: $R$ shows a local peak adjacent to a zero-resistance pocket, while $R_{xy}$ shows a strong enhancement.

\vspace{\baselineskip} 
\noindent\textbf{Superconductivity in twisted \wse}

\noindent
 Fig. 2a shows a higher resolution map of the region enclosed by the white dashed lines in Fig. 1g, highlighting the zero resistance region (dark blue in the chosen colour scale). In Fig. 2b, temperature dependent resistance at a select point in this region (solid circle in Fig. 2a) is compared with the adjacent resistive peak (open circle in Fig. 2a).  The transition to zero resistance shows characteristic behaviour of a two-dimensional superconductor: with decreasing temperature, the resistance exhibits an abrupt downturn followed by a long tail that eventually reaches zero within the measurement noise floor. By comparison, the adjacent resistive peak exhibits an approximately linear in T metallic dependence over the accessible temperature range. Using the empirical convention of identifying the mean field critical temperature to be where $R(T)$ equals 50\% of the normal state resistance gives $Tc=426$~mK at this density and displacement field. 

\begin{figure*}
\includegraphics[width=0.9\linewidth]{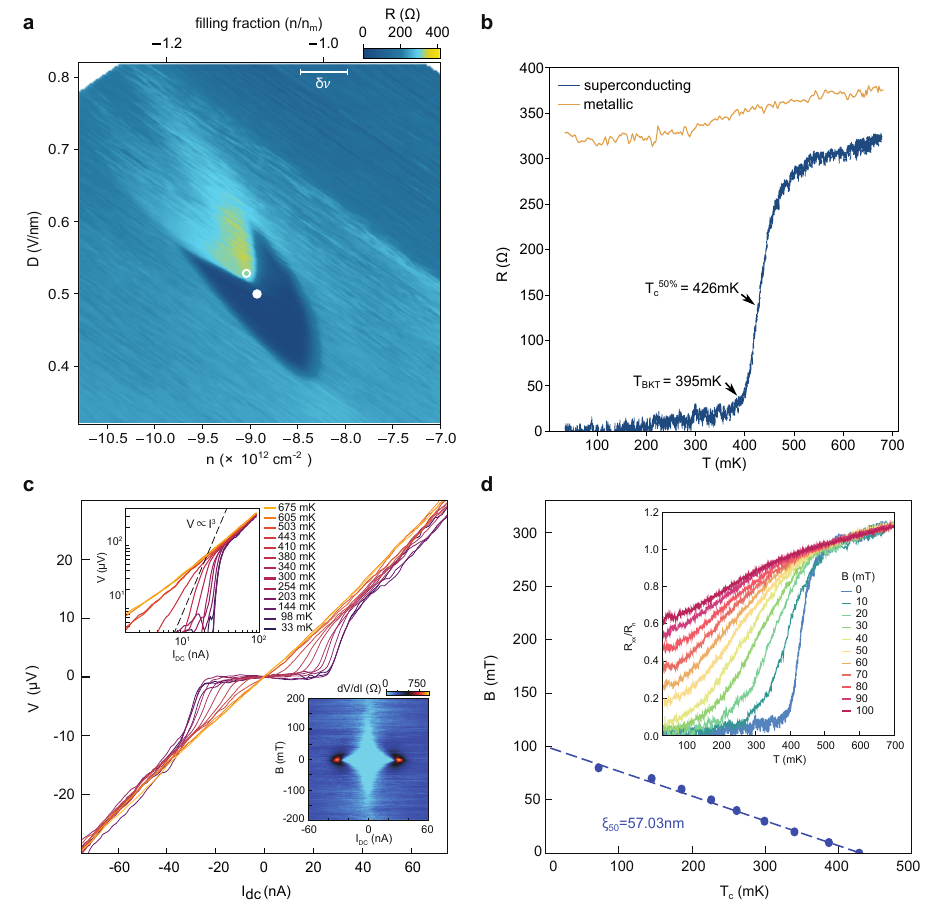}
\caption{{\bf{Superconductivity in twisted \wse.}} 
\textbf{a}, High-resolution map of the region enclosed in the white dashed box in Fig. 1g, showing $R$ around the superconducting pocket. 
Error bar shows the uncertainty in our determination of the filling fraction (see text). Closed circle corresponds to $D=0.5$~V/nm, $n=-8.93$\dens; Open circle corresponds to $D=0.53$~V/nm, $n=-9.05$\dens. \textbf{b} Resistance versus temperature measured at the locations identified in \textbf{a}.  The critical temperature, $T_{C}$, defined at 50\% of the normal-state resistance, and BKT temperature, $T_{BKT}$ determined in \textbf{c} are labelled. \textbf{c}, $V_{xx}-I$ as a function of temperature for the zero resistance location identified in {a}. The top left inset shows the $V_{xx}-I$ curves on a log-log scale. Dashed line indicates where the slope is 3 ($V_{xx} \propto I^{3}$), which gives $T_{\mathrm{BKT}}=395$~mK. The bottom right inset shows d$V_{xx}$/d$I$ versus DC current bias and perpendicular magnetic field \textbf{d}, critical temperature as a function of perpendicular magnetic field. The critical temperature is defined at 50\% of the normal-state resistance from the family of  $R(T)$ curves versus magnetic field shown inset. }
\end{figure*} 

\begin{figure*}
\includegraphics[width=\linewidth]{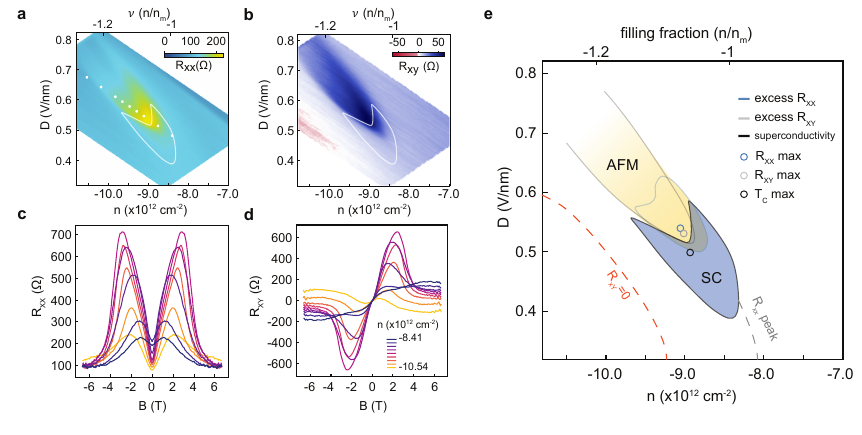}
\caption{{\bf{Magnetism and phase diagram.}} \textbf{a}.\textbf{b}, Longitudinal resistance (\textbf{a}) and Hall resistance (\textbf{b}) as a function of density and displacement field measured at 200~mT T=33mK. A resistive region is seen adjacent to and overlapping with the superconducting phase boundary (indicated by white line) \textbf{c}.\textbf{d}, Longitudinal resistance (\textbf{c}) and Hall resistance (\textbf{d}) versus magnetic field measured at locations indicated by white circles in \textbf{a}, $T=1.6$~K. The legend in \textbf{d} identifies the corresponding densities.     The anomalous magnetotransport is identified with an antiferromagnetic (AFM) order (see text). \textbf{e} Diagram showing the relationship between the superconducting and magnetic regions in the density-displacement field phase space. The AFM phase boundary is identified by the region of excess longitudinal resistance observed in \textbf{a} and excess Hall resistance observed in \textbf{b}. Blue and grey circles mark locations of the maximum value in the 200~mT $R_{xx}$ and $R_{xy}$ maps. The maximum $T_{c}$ position (see Fig. 4) is labelled by the black circle.}
\end{figure*} 

\begin{figure*}
\includegraphics[width=\linewidth]{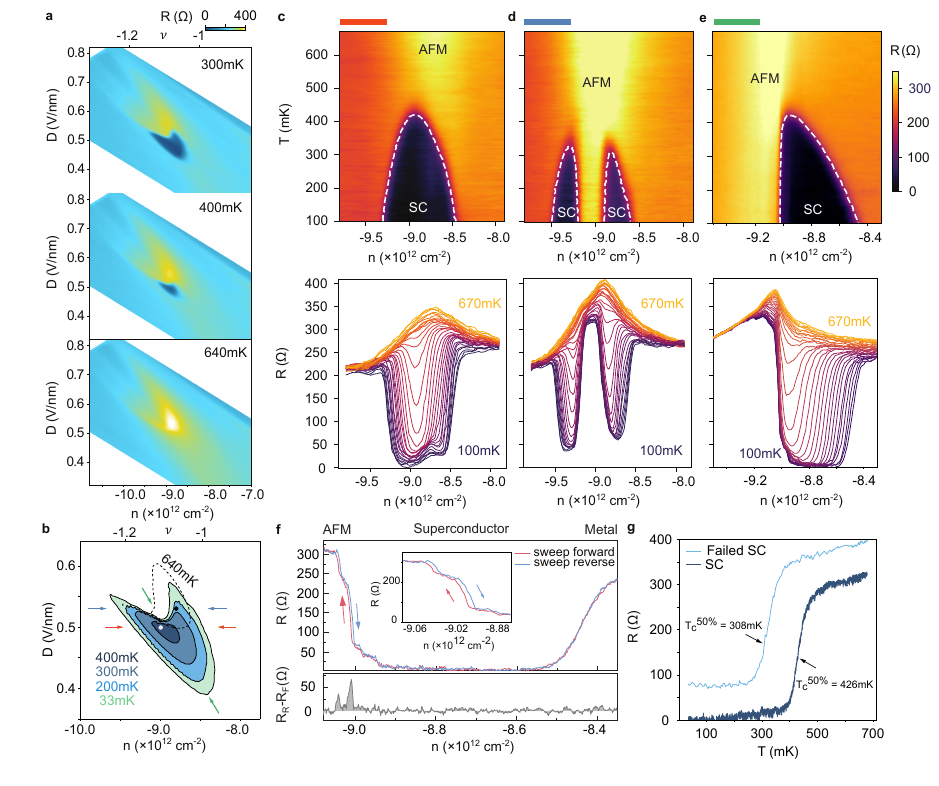}
\caption{{\bf{Temperature dependence of the superconducting boundary.}} \textbf{a}, $R$ versus displacement field and density measured at 300mK, 400mK and 640mK. As the temperature increases the superconducting pocket shrinks. When temperature exceeds $T_{C}$, only the region of excess resistance remains, which we associate with the antiferromagnetic regime. \textbf{b}, Contour plot of the superconducting boundary versus temperature.  The region of excess resistance measured at 640~mK is indicated by the dashed black line.  \textbf{c-e}, Top panels show $R$ plotted versus temperature and density, acquired by sweeping along line cuts in the $D-n$ phase diagram.  The colored bars identify the associated trajectory marked by colored arrows in \textbf{a}.  Red cuts across the superconducting region with largest $T_{C}$ by varying density at fixed displacement field, $D=0.50$~V/nm. Blue cuts across the strongest resistive peak, $D=0.53$~V/n.  Green cuts diagonally through both regions. Lower panels show the associated family of curves.  \textbf{f}.  Forward and reverse trace along the same trajectory measured in \textbf{e} showing evidence of hysteresis (emphasized inset) at the SC-AFM transition, and no hysteresis at the SC-normal metal boundary. Bottom panel plots the difference between the traces. \textbf{g}, Comparison of $R(T)$ measured in the region of overlap between the AFM and SC boundaries (black circle in \textbf{b}), and outside the overlap region (white circle in \textbf{b}).  The light blue curve, which is measured in the overlap region, shows characteristics of a superconducting transition but saturates to finite value before reaching zero resistance. }
\end{figure*} 

Fig. 2c shows the temperature dependent voltage-current ($V-I$) response for this same point.  In the low-temperature limit, non-linear response typical of superconductivity is observed.  We estimate the critical current at base temperature to be  approximately $I_{c}=30~nA$.  By fitting the power-law dependence (Fig. 2c inset left), we extract a Berezinskii–Kosterlitz–Thouless (BKT) transition temperature of $T_{\mathrm{BKT}} = 395$ mK (see methods) which is close to the estimated mean field temperature ($T_{\mathrm{BKT}}/T_{c}\sim0.93$), as expected for a low-disorder superconductor \cite{beasleyPossibilityVortexAntivortexPair1979}. 
Inset right in Fig. 2c is a plot of $dV/dI$ versus DC current  and magnetic field.  The qualitative behaviour is again consistent with superconductivity and indicates a base temperature critical field of $B_{c\perp}\approx100$~mT.

Fig. 2d shows the B-T phase diagram, determined using the same 50\%$R_{n}$ definition for T$_{c}$ (a family of normalized $R(T)/R_{n}$ versus $B$ curves are shown inset in Fig. 2d), where the normal state resistance was defined by the knee of the transition. 
The critical field varies linearly with temperature, consistent with the Ginzburg-Landau equation,  $H_{c2}^{\perp} = \phi_{0}/(2\pi\xi(0)^2)(1 - (T/T_c))$, where $\phi_{0}$ is the flux quantum, $T_{c}$ is the zero field critical temperature and $\xi(0)$ is the coherence length.  The slope of the fitted line gives  $\xi(0) \approx 57$ nm.  
The normal-state square resistance, defined at the knee of the transition, is $R=150$~$\Omega$. The mean free path, $l_{m}$, can be estimated From the Drude model,  
$R = h/{e^{2}k_Fl_{m}}$,  where $h$ is Planck’s constant, $e$ is the elementary charge, and $k_F= \sqrt{2\pi n}$,  giving  $l_{m}=229$~nm.  This in turn gives an estimate for the disorder ratio to be $d=\xi(0)/l_{m}\sim0.25$, suggesting that the superconductor is in the clean limit.


\vspace{\baselineskip}
\noindent\textbf{Interplay between superconductivity and magnetic order}

\noindent
To further examine the resistive feature adjacent to the superconducting phase, we suppress the superconductor by applying an out-of-plane magnetic field. Fig. 3a,b shows $R_{xx}$ and $R_{xy}$  measured at $B_{\perp}= 200$~mT, with the white line indicating the boundary of the superconducting pocket at zero field. In both channels we observe an excess resistance over a finite range of density and displacement field.  Interestingly, the boundaries of the resistive and superconductor features occupy predominately separate regions,  with a small overlap. 

Fig. 3c,d shows $R_{xx}$ and $R_{xy}$ versus $B_{\perp}$, measured along the locations indicated by white circles in Fig. 3a. The magnetoresistance characteristics in our 5$^{o}$ device are very similar to those reported in  recent studies on 4.2$^{o}$ t\wse\cite{ghiottoStonerInstabilitiesIsing2024}, albeit with lower characteristic temperature (see SI).  The peak in the Hall resistance and noticeable but less dramatic peak in the longitudinal resistance was interpreted to arise from a Fermi surface reconstruction, most likely due to antiferromagnetic (AFM) order induced by the large susceptibility occurring at wave vectors that connect the Van Hove points of the spin up and down Fermi surfaces. The antiferromagnetic order partially gaps the Fermi surface. In the supplemental information we calculate the density of states and wave vector-dependent susceptibility as a function of carrier concentration and displacement field, using the highest energy band from \cite{crepelBridgingSmallLarge2024}. The magnitude of the susceptibility increases with increasing displacement field, but the Fermi surface becomes less nested. Based on this trend we suggest that above a critical displacement field  the Stoner criterion is satisfied, producing magnetic order, and that in the vicinity of the magnetic state, superconductivity is stabilized. 


A summary schematic of the full AFM-superconductor phase diagram is shown in Fig. 3e.  Yellow indicates the region of AFM ordering identified by contours of excess resistance in $R_{xx}$ (blue line) and $R_{xy}$ (grey line).  The blue region represents the arrow shaped superconducting pocket observed at base temperature (Fig. 2a).  The dashed grey and red dashed lines indicate the $R_{xx}$ peak and $R_{xy}=0$ trajectories associated with the Van Hove singularity (Fig. 1g,h). 
The locations of maximal excess resistance in $R_{xx}$ (blue circle) and $R_{xy}$ (grey circle) are essentially coincident with each other, and are both adjacent to but not apparently overlapping the location of peak superconducting $T_{c}$ (black circle, see discussion of Fig. 4 for more details).  We note that the appearance of superconductivity near an interaction-driven phase transition in the spin-valley ordering is phenomenologically similar to observations in the graphene-based flat band systems\cite{caoUnconventionalSuperconductivityMagicangle2018,luSuperconductorsOrbitalMagnets2019,liuTuningElectronCorrelation2021,parkTunableStronglyCoupled2021,haoElectricFieldTunable2021,zhouIsospinMagnetismSpinpolarized2022,zhouSuperconductivityRhombohedralTrilayer2021,zhangEnhancedSuperconductivitySpin2023,liTunableSuperconductivityElectron2024} which could point towards a universal mechanism.


We next examine how the phase diagram evolves with temperature. Fig.4a shows plots of $R$ versus displacement field and density, measured at temperatures 300 mK, 400 mK and 640 mK. As temperature increases, the boundary of the superconducting pocket (dark blue) gradually shrinks. 
At 640 mK,  we see only a region of excess resistance remaining (yellow in the color scale),  corresponding to the same phase we identified when suppressing the superconductor with magnetic field (Fig. 3a). 
Fig. 4b shows a contour plot of the superconducting boundary at different temperatures (defined by 150 $\Omega$, about 50\% of the normal state resistance). The dashed black lines mark a contour boundary of the excess resistance  corresponding to $R = 350~\Omega$ in the 640 mK map. It is notable that as temperature increases, the boundary between the magnetic and superconducting phases remains sharp, whereas the boundary furthest from the magnetic phase gradually recedes.  This indicates that the region of most robust superconductivity is directly adjacent to the magnetic phase, but also reveals that the phase transition between to the magnetic and superconducting states is sharply defined - possibly first-order.

Figs. 4c-e show cross-sections through the full temperature-displacement field-density phase space obtained by measuring temperature dependence along the trajectories marked by coloured arrows in Fig. 4b. In each we get a sense of how the superconducting dome interplays with the surrounding phases.  Fig. 4c, obtained by varying density at fixed $D=0.5$~V/nm, cuts across the point of strongest $T_{c}$ and intersects only the weak edge of the magnetic phase (red arrows in Fig. 4b).  The corresponding superconducting dome is approximately symmetric about the maximum $T_{c}$. However, the line cuts reveal that the resistance does not fully reach zero in the region of overlap with the magnetic phase. Fig. 4d again corresponds to fixed $D$ but this time cutting across the region of strongest magnetic signature (blue arrows in Fig. 4b). Here the apparent AFM state divides the dome in half, persisting all the way to base temperature.  Finally, Fig. 4e cuts across the long diagonal pocket, intersecting the strongest superconductor and magnetic phase regions.  The superconducting $T_{c}$ shows a nearly vertical drop to zero when it encounters the AFM phase boundary, but shows a more usual gradual dome-like shape when transitioning to the normal metal on the other side. 

Fig. 4f shows a single forward and reverse gate sweep measured along along this same trajectory.  We observe a weak but measurable hysteresis at the SC-AFM transition, and no evidence of hysteresis across the SC-metal transition under identical conditions. This provides further evidence of a first-order phase transition across the SC-AFM boundary, and reinforces our interpretation that the Fermi surface reconstruction results from AFM ordering\cite{yangDomainwallSuperconductivitySuperconductor2004}.


Finally, in Fig. 4g, we compare $R(T)$ between regions where the superconductor and AFM overlap versus where they do not overlap, corresponding to the positions marked by the white and black circles, respectively, in Fig. 4b. In the overlapping region (light blue curve), $R(T)$ shows a sharp transition that closely resembles the non-overlapping region (dark blue curve), but with lower $T_{c}$.  Interestingly the transition in the overlap region does not reach zero resistance but instead saturates to a finite value. This behaviour suggests a failed superconductor, which we conjecture arises from a mixed phase of AFM and superconducting regions.

\vspace{\baselineskip}
\noindent\textbf{Discussion and conclusions}

\noindent 
Having demonstrated robust superconductivity in \twse, we next consider what its nature might be. We first consider the pairing strength. Estimating the Fermi energy to be $E_{F}\sim50$~meV ($\sim$half of the full band width) gives the ratio for the critical temperature, $T_{C}$, versus Fermi temperature, $T_{F}=E_{F}/k_{B}$,  to be $T_{c}/T_{F}\sim0.0008$, a value typically associated with the weak pairing 
 ``BCS'' regime\cite{bardeenTheorySuperconductivity1957}. The BKT relation implies that at $T=T_{BKT}$ the superfluid stiffness, $\rho_{s}(T_{c})=\frac{2}{\pi}T_c$ is approximately 0.25~K, about two orders of magnitude  less than the clean-limit superfluid stiffness $\sim 20K$ calculated from the band structure given in \cite{crepelBridgingSmallLarge2024}. 
 The small value of the superfluid stiffness, large value of the coherence length relative to the \moire lattice constant ($\xi_0/a_m\sim 15$) and large bandwidth all support BCS pairing.

It has been theoretically proposed that phonon-mediated interactions could induce a conventional type superconductivity in the graphene flatband systems,  \cite{balentsSuperconductivityStrongCorrelations2020, wuTheoryPhononMediatedSuperconductivity2018, liuTuningElectronCorrelation2021}, with similar possibility anticipated in for  \twse\cite{wuHubbardModelPhysics2018, hsuSpinvalleyLockedInstabilities2021}.  Although we estimate that this system is in in the BCS limit, we nonetheless conjecture that the superconductivity is unconventional in the sense of being induced by a non-phonon mechanism. A feature that distinguishes our system from the twisted multilayer graphene systems \cite{caoUnconventionalSuperconductivityMagicangle2018,luSuperconductorsOrbitalMagnets2019,liuTuningElectronCorrelation2021,parkTunableStronglyCoupled2021,haoElectricFieldTunable2021}, but finds similarity with rhombohedral trilayer graphene\cite{zhouSuperconductivityRhombohedralTrilayer2021} and Bernal bilayer graphene \cite{zhouIsospinMagnetismSpinpolarized2022,zhangEnhancedSuperconductivitySpin2023,liTunableSuperconductivityElectron2024}, is  the emergence of superconductivity coincident with a Fermi-surface reconstruction where magnetic ordering in the spin-valley sector is also evident. In twisted graphene there is ambiguity since the narrowness of the band makes it difficult to determine proximity relationships between the various ordered states\cite{balentsSuperconductivityStrongCorrelations2020}. If the superconductivity in our device were phonon mediated with an enhancement by the Van Hove density of states, we might expect it to track the VHs throughout the band. That instead we observe superconductivity only within a narrow window, concomitant with magnetic ordering, points toward the magnetic order playing a critical role such as mediating pairing via spin fluctuations\cite{fischerSpinfluctuationinducedPairingTwisted2021,mathurMagneticallyMediatedSuperconductivity1998,Zhai09,kleblCompetitionDensityWaves2023}.  

In summary our interpretation is that at zero displacement field, correlations are too weak for superconductivity to be observable within experimental temperature.  However, under application of a displacement field, the resulting increase in density of states at the VHs, and increasing magnetic susceptibility, reach a threshold where a Fermi surface reconstruction to an antiferromagnetic state results.  Spin fluctuations in the vicinity of the AFM phase are then responsible for mediating the superconducting pairing state. This interpretation would explain the absence of superconductivity everywhere except in proximity to the AFM phase, while also being qualitatively consistent with the superconducting phase defined by a $T_{c}$ that rises on approaching the AFM phase, reaches a peak value at the AFM boundary, and then abruptly disappears.  Finally we note that this behaviour is very similar to antiferromagnetic heavy Fermion systems in the clean limit where spin fluctuations provide the pairing mechanism\cite{mathurMagneticallyMediatedSuperconductivity1998,scalapinoCommonThreadPairing2012}.

A full understanding of the superconductivity that we observe in \twse will certainly require further experimental and theoretical work.  In particular a more comprehensive understanding of the interplay between the magnetic order and superconducting state, beyond what can be accomplished by transport alone,  will be necessary to determine whether the two phases are cooperative, competing, coincident or some combination of all three.  The ability to move throughout the magnetic-superconductor phase diagram within a single device via purely electrostatic gating offers unprecedented opportunity to identify universal versus material specific characteristics of this interplay and could provide new insight beyond what has been achieved in the study of conventional high $T_{c}$ and heavy Fermion systems. A key question to address will be how the phase diagram evolves with twist angle. This also provides a unique opportunity compared with the graphene systems where it is already established that in the TMDs, correlations survive and evolve over several degrees of twist angle, compared to the relatively narrow ``magic'' windows in twisted multilayer graphene, and the essentially singular parameter space of rhombohedral stacked multilayers.

\section{Acknowledgments}
\begin{acknowledgments}
This research is primarily supported by US Department of Energy (DE-SC0019443). WSe2 was synthesized by JH and KM under the support of the Columbia University Materials Science and Engineering Research Center (MRSEC), through NSF grants DMR-2011738. D.M. and M.C. acknowledge support from the Gordon and Betty Moore Foundation’s EPiQS Initiative, Grant GBMF9069. K.W. and T.T. acknowledge support from the JSPS KAKENHI (Grant Numbers 21H05233 and 23H02052) and World Premier International Research Center Initiative (WPI), MEXT, Japan. J.H. and C.R.D. acknowledge additional support from the Gordon and Betty Moore Foundation’s EPiQS Initiative, Grant GBMF10277. 
\end{acknowledgments}

\section{Author Contributions}
Y.G. fabricated the device. Y.G., J.P., J.S. and C.R.D. performed the electronic transport measurements and analyzed the data. L.H. grew the \wse crystals under the supervision of J.H. and K.B; M.C. grew the \rucl crystals under the supervision of D.G.M.; K.W. and T.T. grew the hexagonal boron nitride crystals. A.M. performed theoretical modeling. Y.G., A.M., A.P. and C.R.D. wrote the manuscript with input from all authors.

\section*{Competing financial interests}
The authors declare no competing financial interests.

\bibliography{yg_full}

\section{Methods}

\subsection{Device fabrication}

We use a thin polycarbonate film to pick up the exfoliated layers from \sio with the dry transfer method \cite{wangOneDimensionalElectricalContact2013}. We use thin boron nitride as the dielectric spacer to increase the displacement field and density range in the phase diagram. Few layer graphites are used for top and bottom gates. We use few-layer graphite as contacts to the \wse layers and a thin RuCl$_3$ flake is stacked in the contact region to improve contacts to the \wse layers. We stack the layers from top to bottom in the sequence of: topmost bn ($\sim$11nm) -graphite top gate - top bn ($\sim$5nm) - graphite contacts - twisted \wse layers - \rucl layer - bottom bn ($\sim$17nm) - graphite bottom gate. The alignment of the graphite top gate, graphite contact and \rucl is especially taken care of to make sure that the top gate is local to the channel region and avoid overlapping with the contact region. For stacking the twisted \wse, we pre-cut a \wse monolayer into two pieces with an AFM tip. We pick up one of the pieces and rotate the transfer stage by about 5 degree before picking up the second piece. Finally we drop down the stack onto a \sio substrate. 

A metal contact gate is deposited on top of the stack to smooth out the gap between the heavily doped contact region by \rucl and the intrinsic channel region. Metal leads to the gates and the graphite contacts are deposited after etching through the bn with the same PMMA mask. Finally, the device is etched into a Hall bar shape with alternating SF$_{6}$ and O$_{2}$ plasmas with a 30W O$_{2}$ plasma at the end to etch through the \rucl flake. PMMA is left on the chip after the etching process to avoid any exposure of \rucl to solvents. The device is stacked and fabricated in the same way as in this reference\cite{packChargetransferContactHighMobility2023}.  See supplementary material for device image. 

\subsection{Measurements}
Transport measurements were performed in a variable temperature cryostat with a base temperature of 1.6 K and in a dilution fridge with a base temperature of 33 mK. Four and two terminal resistance measurements were carried out using a low-frequency lock-in technique at frequencies ranging from 6 Hz-25 Hz. Characterization for  superconducting region is done with a current source bias of 5nA. IV characteristics were measured in DC using a sourcemeter and nanovoltmeters. dV/dI measurements were done using lockin SR860 to provide an AC signal on top of the DC signal. The current signal is amplified using a SR570 and measured with SR860s and a nanovoltmeter. 

\subsection{BKT temperature extraction}
We plot $V_{xx}-I$ curves measured at different temperatures on a log-log scale (Fig. 2c inset). We extract the power at different temperatures in the transition region.  To find the temperature where the curve follows $V \propto I^3$, we plot power versus temperature and interpolate to find where it crosses 3. (See SI)

\subsection{Theory}
We used the formalism of Ref.~\cite{crepelBridgingSmallLarge2024} which presents a three band tight binding model with parameters fit to DFT calculations  to determine the band structure. This paper gives parameters only for angles up to $4.5^\circ$; we extrapolated  to $5^\circ$, in the notations of Ref.~\cite{crepelBridgingSmallLarge2024} we find (all energies in meV) $\delta = 15$, $t_1^{th} = 15$, $t_1^{hh} = 2$ $t_2^{th} = 0$, $t_1^{tt} = -2$, $t_3^{hh} = 4.$ and we considered displacement fields up to $18meV$. We computed the band energies on a dense grid of $k$-points in the Brillouin zone, identified the highest lying electron band $\varepsilon^3_k$ and then used these energies to construct the single-spin density of states $N_0(\omega)=\int_{BZ}\frac{d^2k}{(2\pi)^2}\delta(\omega-\varepsilon^3_k)$ and the transverse $\chi^\pm(q,0)=\mathcal{P}\int_{BZ}\frac{d^2k}{(2\pi)^2}\frac{f(\varepsilon^{3,\uparrow}_k)-f(\varepsilon^{3,\downarrow}_{k+q})}{\varepsilon^{3,\uparrow}_k-\varepsilon^{3,\downarrow}_{k+q}}$
 and longitudinal $\chi^{zz}(q,0)=\mathcal{P}\sum_\sigma\int_{BZ}\frac{d^2k}{(2\pi)^2}\frac{f(\varepsilon^{3,\sigma}_k)-f(\varepsilon^{3,\sigma}_{k+q})}{\varepsilon^{3,\sigma}_k-\varepsilon^{3,\sigma}_{k+q}}$ susceptibilities ($\mathcal{P}$ denotes principal value). At this level of approximation there is no qualitative difference between the longitudinal and transverse susceptibilities.  Fig SI. 7 show the density of states for different displacement fields.
\section{Data availability}
The data that support the plots within this paper and other findings of this study are available from the corresponding author upon reasonable request.

\end{document}